\documentclass[lettersize,journal]{IEEEtran}
\IEEEoverridecommandlockouts
\usepackage{graphicx} 
\usepackage{amssymb, amsmath, amsfonts, amsthm}
\usepackage{algorithmic}
\usepackage{multirow}
\usepackage{bm}
\usepackage{nomencl}
\usepackage{mcite}
\usepackage[noadjust]{cite}
\usepackage{enumitem}
\usepackage{pstricks}
\usepackage{textcomp}
\usepackage{xcolor}
\usepackage{array}
\usepackage{booktabs}
\usepackage{flushend}
\usepackage{url}

\def\BibTeX{{\rm B\kern-.05em{\sc i\kern-.025em b}\kern-.08em
    T\kern-.1667em\lower.7ex\hbox{E}\kern-.125emX}}

\begin{document}

\allowdisplaybreaks
\title{
Resource-Aware Rolling-Horizon Controller for Renewable-Based Virtual Power Plants Providing Automatic Frequency Restoration Reserves
\thanks{This work is financed by PREDFLEX-CM (TEC-2024/ECO-287), funded by Community of Madrid, and PLANTAFLEXILIAR (PID2025-174473OB-I00), funded by MICIU/AEI/10.13039/501100011033 and FSE+. 
}}

\author{\IEEEauthorblockN{Marco Vinicio Avenda\~{n}o-Caiza, Juan Diego Rios-Pe\~{n}aloza, Javier Rold\'{a}n-Pérez, \textit{Senior Member, IEEE}, \\ Jos\'{e} Luis Rodr\'{i}guez-Amenedo, and Milan Prodanovi\'{c}, \textit{Senior Member, IEEE}} \vspace{-0.2cm}
}
\maketitle

\begin{abstract}
Virtual power plants (VPPs) are an effective solution for increasing the penetration of renewable energy sources (RES) in power systems. Moreover, VPPs can provide ancillary services if generators and loads are closely connected and properly coordinated.
This aspect has been addressed in several studies, yet they mostly focus on the dispatch optimisation level, leaving real-time operation aspects unresolved.
To close this gap, a central controller for VPPs based on a rolling-horizon optimisation is proposed in this work.
Its main objective is to maximise revenues while delivering frequency restoration services.
A day-ahead optimisation firstly defines the power setpoint and the upward and downward reserves of the VPP.
Then, the rolling-horizon optimiser, executed every minute, distributes the power and reserve references between the VPP units to achieve their close tracking while taking into account resource availability (wind speed, irradiance, etc.),  operational constraints and battery degradation. 
The effectiveness of the controller is tested using a two-area interconnected power system under different RES generation and demand profiles. 
The simulations are performed in MATLAB/Simulink and include detailed dynamic models of the VPP elements, demonstrating the applicability of the algorithm to real systems.
The day-ahead and the rolling-horizon optimisation problems are modelled using YALMIP and solved using Gurobi.
The obtained results demonstrate the VPP supporting grid frequency restoration by optimally allocating its resources, even under the presence of forecast uncertainty. 
\end{abstract}
\vspace{-0.2cm}
\vspace{-0.5cm}

\section*{List of Variables}
\vspace{-0.2cm}
\noindent \textit{General}\par
\vspace{+0.2cm}
{\renewcommand{\arraystretch}{1.1}
\noindent\begin{tabular}{@{}p{0.28\columnwidth}@{\hspace{6pt}}p{0.69\columnwidth}@{}}
$\tilde{\cdot}$, $\hat{\cdot}$ & Forecasted value, decision variable. \\
$k$, $\Delta t$ & Optimisation step, sampling period. \\
$B$, $p^*_{SC}$ & Frequency bias factor and frequency-restoration reference. \\
$Set$ & Binary signal selecting the reserve mode. \\
$P^{nom}_{\{w,PV,bat\}}$ & Rated power of the units. \\
$\tilde{p}_{load}$ & VPP internal load forecast. \\
$\tilde{v}_{w}$, $\tilde{G}$ & Wind speed and irradiance forecasts. \\
$n_w$, $A_p$, $\rho$, $c_p$ & Turbines per farm, blade swept area, air density, power coefficient. \\
$SOC$, $\eta$, $Q_n$ & BESS state of charge, efficiency and nominal capacity. \\
\end{tabular}}

\vspace{+0.1cm}
\noindent \textit{Day-Ahead Optimisation}\par
\vspace{+0.05cm}
{\renewcommand{\arraystretch}{1.1}
\noindent\begin{tabular}{@{}p{0.28\columnwidth}@{\hspace{6pt}}p{0.69\columnwidth}@{}}
$\pi_1$, $\pi_2^{up,down}$ & Energy and reserve capacity prices. \\
$p^*_{VPP}$, $R^{*\;up,down}_{VPP}$ & Scheduled power and reserve commitments. \\
\end{tabular}}

\vspace{+0.2cm}
\noindent \textit{Cost Functions}\par
\vspace{+0.2cm}
{\renewcommand{\arraystretch}{1.1}
\noindent\begin{tabular}{@{}p{0.28\columnwidth}@{\hspace{6pt}}p{0.69\columnwidth}@{}}
$C_{op,\{w,PV\}}$ & Operating costs of wind and PV. \\
$C_{ws}$ & Wind-speed-dependent degradation cost. \\
$C_{dis}$, $C_{ch}$ & BESS degradation costs. \\
$C_{\Delta p}$, $C^{up,down}_{\Delta R}$ & Penalties on power and reserve tracking errors. \\
$(\cdot)^{ct,1}$, $(\cdot)^{ct,2}$ & Breakpoints of the piecewise cost curves. \\
$m_{ws}$, $m_{op,\{w,PV\}}$, $m_{\{dis,ch\}}$ & Slopes of the corresponding cost curves. \\
$a_1$, $a_2$ & Power and reserve penalty factors. \\
$\Delta\hat{p}_{VPP}$, $\Delta\hat{R}^{up,down}_{VPP}$ & Power and reserve tracking errors. \\
\end{tabular}}

\vspace{+0.2cm}
\noindent \textit{Rolling-Horizon Optimisation}\par
\vspace{+0.2cm}
{\renewcommand{\arraystretch}{1.1}
\noindent\begin{tabular}{@{}p{0.28\columnwidth}@{\hspace{6pt}}p{0.69\columnwidth}@{}}
$\hat{p}_{\{VPP,w,PV,bat\}}$ & VPP, wind generators, PV plant and BESS active power. \\
$\hat{p}_{b,\{dis,ch\}}$, $\hat{p}_{bat}$ & BESS discharging, charging and operating power. \\
$\tilde{p}^{\max}_{\{w,PV\}}$, $P^{\max}_{bat}$ & Maximum available power of each unit. \\
$\bm{\hat{K}_{op}}$ & Operating factors of the renewable units. \\
$\hat{R}^{up,down}_{j}$, $\hat{R}^{up,down}_{VPP}$ & Reserve capacity of each unit and of the VPP. \\
$\bm{f^{up,down}_a}$ & Reserve availability factors. \\
\end{tabular}}

\section{Introduction}
The provision of ancillary grid services is essential for maintaining power-system stability and has become increasingly important with the soaring numbers of converter-interfaced generators, mainly connecting renewable energy sources (RESs).
Aggregation of these sources allows a more efficient provision of ancillary services.
Virtual power plants (VPPs) concept represents an attractive way for aggregation and coordination of geographically distributed energy resources, including flexible loads, generation units, and energy storage systems (ESSs).
These units are coordinated via a central controller to operate as a single entity, yielding advantages in terms of flexibility and operational efficiency.
Several recent studies have explored and proposed new VPP capabilities, like coordinated operation to follow a reference (using optimisation and participation in electricity markets) and the provision of ancillary services~\cite{VPPs_Models_and_Markets}.
The role of VPPs in electricity markets is currently receiving increased attention, particularly for the provision of ancillary services~\cite{ZARE2026116448,GHOLAMI2025101959}, although their ability to deliver specific services depends on the technical characteristics of the aggregated resources, network constraints, and market regulations~\cite{VPPs_concepts}.
Indeed, because of the dynamic characteristics of energy resources, VPPs exhibit complex responses that are challenging to manage.
This challenge is even more highlighted by the increasingly demanding requirements imposed on the provision of ancillary services in modern power systems.
In this scenario, the application of optimisation frameworks that account for dynamic properties and resource availability (wind, storage, etc.) is essential.

In line with VPPs evolution towards more active participation in power system operation, a substantial body of research has addressed the scheduling of their resources for providing energy and ancillary services.
The day-ahead participation of VPPs in energy and reserve markets has been studied considering the uncertainty of resource availability, market prices, and reserve deployment requests~\cite{VPP_DAM}, also through robust formulations~\cite{VPP_Robust_Scheduling}. 
Some works optimised the joint scheduling of VPPs, including battery degradation costs~\cite{Optimisation_Models_DAM}.
These degradation costs have also been considered for the participation in frequency regulation markets~\cite{Optimal_BESS_Participation}.
All the works mentioned before, however, only determine the scheduling profiles but do not address the real-time operation of the VPP, where the actual availability of the resources and the activation of the reserves affect its performance.

There have been more recent studies addressing the real-time coordination of VPPs.
For example, Häberle \textit{et al.}~\cite{DVPP_Control_Design} proposed a hierarchical control architecture based on adaptive dynamic participation matrices, enabling the provision of fast ancillary grid services in heterogeneous power systems.
Other works enabled the direct participation of VPPs in secondary frequency control through a real-time power redispatch~\cite{Direct_Part_DVPP_in_FC}, incorporated VPPs into load frequency control through a decentralised control strategy~\cite{Load_Freq_Control_Wang}, and considered cyber-physical uncertainties through a robust formulation~\cite{GUO202593}. 
In some cases, the regulation signal is distributed between the resources according to their dynamic characteristics and available capacity~\cite{Cord_Freq_Reg_Oshnoei}.
Also, some works determined the power reserve required by a VPP and allocated it to the inverter-based resources according to their energy limitations and economic characteristics~\cite{DVPP_Robust_Freq_regul, Min_Reserve_and_Allocation_Zhu}.
Although these approaches enhance the operation and robustness of VPPs, they generally assume predefined reserves or regulation commands and focus on their allocation or tracking, rather than on their continuous techno-economic optimisation.
In this direction, rolling-horizon optimisation and model predictive control approaches have been developed to update operational decisions as new measurements and forecasts become available. 
For example, Bolzoni \textit{et al.}~\cite{Optimal_VPP} complemented the day-ahead schedule of multiple grid-support services with a model predictive controller that reduces imbalance costs during operation.
Amini \textit{et al.}~\cite{MPC_Amini} dispatched VPPs according to their energy state, and Gulotta \textit{et al.}~\cite{Gulotta} proposed a real-time management of the reserves under short-term uncertainty.
Zhuang \textit{et al.}~\cite{SFC_of_Dist_Level_VPP_Zhuang} made use of stochastic model predictive control in a distribution-level VPP to provide reserves under PV uncertainty and storage capacity limits.
Chen \textit{et al.}~\cite{Real_time_op_VPPs_Chen} accounted for the temporal coupling among resources to optimally allocate regulation commands, giving priority to the least-cost units.
Sun \textit{et al.}~\cite{Decentralized_FR_for_VPPs} proposed a method based on the game theory to share the frequency regulation service among the resources available in the VPP.
Srivastava \textit{et al.}~\cite{Enabling_DER_Freq_reg_Markets} developed a hierarchical framework for participation in frequency-regulation markets and distributing the regulation signal between the resources.
These formulations improve the economic utilisation of the available flexibility. 
However, they rely on simplified models and, therefore, do not guarantee the feasibility of the resulting decisions under realistic scenarios. 

Overall, the existing literature distinguishes between optimisation-oriented and control-oriented approaches.
Market and scheduling studies provide detailed formulations for maximisation of revenues, uncertainty management, and energy and reserve allocation, but these usually use significant simplifications of the dynamics of the VPP and the interconnected grids.
Conversely, dynamic-control studies demonstrate the capability of VPPs to provide primary or secondary frequency support, but commonly assume predetermined reserve levels, power commands, or participation factors.
Real-time disaggregation and model-predictive approaches partially bridge these perspectives~\cite{10764748}, although the simultaneous consideration of renewable-resource availability, market commitments, battery degradation, and operational restrictions remains insufficiently explored.
This work addresses this gap by proposing a centralised controller based on a rolling-horizon optimisation framework that jointly considers economic benefits and technical performance.
The controller aims to closely follow the energy and reserve commitments established in the day-ahead market by an offline optimisation while providing automatic frequency restoration reserve in a multi-area power system.
The controller takes into consideration the real-time operational aspects of the VPP units.
It is designed to maximise the technical and economic performance by optimising the utilisation of available resources, while avoiding operation under unsuitable conditions.
The effectiveness of the proposed controller is tested in a two-area power system.
The VPP, constituted by two wind and one PV power plants, one energy storage system, and distributed loads, is connected to one of the areas.
Real production and demand profiles are used, and the performance of the controller is assessed when the forecast production and demand profiles differ from the real ones.

The remainder of this work is explained in the following lines.
In Section~\ref{sec.overview}, the VPP and its controller are explained, together with the optimisation methodology. 
The details of the VPP elements, including their control system, are presented in Section~\ref{sec.power.sys.model}. 
In Section~\ref{sec.proposed.optimiser}, the VPP optimiser is described, and the modelling equations are explained in detail.
The simulation results are described and discussed in Section~\ref{sec.results}. 
The paper ends with Section~\ref{sec.conclusion}, where the main conclusions are presented.
\vspace{-0.1cm}
\section{Overview}
\label{sec.overview}
\begin{figure}[!t]
\centering
\vspace{-0.1cm}
\includegraphics[width=0.98\columnwidth]{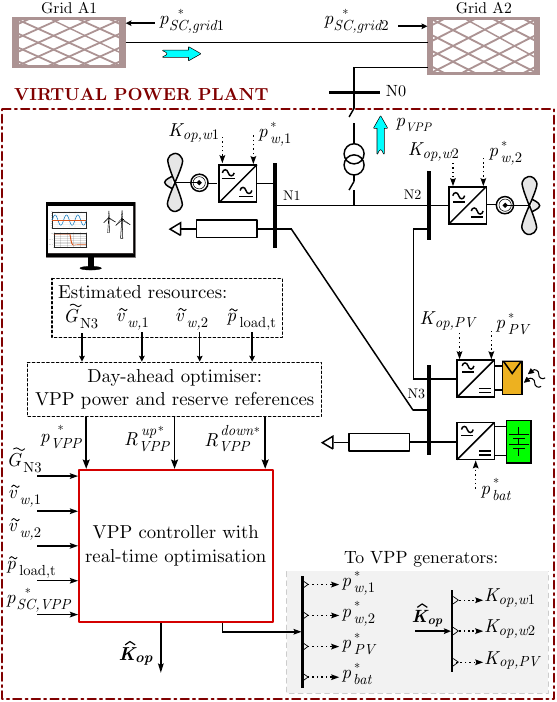}
\vspace{-0.1cm}
\caption{Electrical diagram of the system, including two transmission areas and a VPP providing frequency restoration support.}
\vspace{-0.4cm}
\label{fig:VPP_Single_Diagram}
\end{figure}
\subsection{System Description}
The electrical diagram of the system studied in this work is shown in Fig.~\ref{fig:VPP_Single_Diagram}. 
It comprises two interconnected transmission areas and a VPP connected to one of them.
The system rated frequency and voltage are 50~Hz and 220~kV; the rated powers of Grids A1 and A2 are 150~MW and 100~MW, respectively.
The VPP operates at a rated voltage of 100~kV and comprises two 60~MW wind power plants connected to nodes N1 and N2, a 60~MW solar PV power plant connected to node N3, and a 60~MW battery energy storage system (BESS), also connected to node N3.
In addition, two variable loads connected to nodes N1 and N3 represent the internal power consumption of the VPP.
The wind and solar power plants are integrated into the grid using grid-following converters, while the BESS is integrated using a grid-forming converter.

The block diagram representing the VPP controller is shown in 
Fig.~\ref{fig:VPP_Single_Diagram}. 
A day-ahead optimisation determines the power ($p^*_{VPP}$), and upward ($R^{*\;up}_{VPP}$) and downward ($R^{*\;down}_{VPP}$) reserve setpoints of the VPP.
During real-time operation, a system-level secondary controller determines the frequency-restoration contribution of grid A1 ($p^*_{SC,grid1}$), A2 ($p^*_{SC,grid2})$, and the VPP $(p^*_{SC,VPP})$.
This controller keeps the system frequency at its nominal value and ensures accurate power exchanges, following standard secondary (AGC-type) control practice~\cite{Kundur_book1}.
As shown in Fig.~\ref{fig:Secondary_Controller}(a), each branch of this controller applies a frequency bias factor ($B_{g1}$, $B_{g2}$, and $B_{VPP}$), and computes the power variations to provide frequency restoration support. 
Then, PI controllers compute the reference powers sent to the local controllers of areas A1, A2, and the VPP. 
Finally, the real-time controller of the VPP is based on a rolling-horizon optimisation problem and constitutes the main contribution of this work.
The allocation of the VPP reference within its internal sources is made according to the availability factors of each source, which depend on availability and key technical parameters of the individual units.
These factors are provided by the rolling horizon optimiser proposed in this work.
\subsection{Optimisation Framework and Virtual Power Plant Controller Overview}
\begin{figure}[!t]
\centering
\vspace{-0.1cm}
\includegraphics[width=0.98\columnwidth]{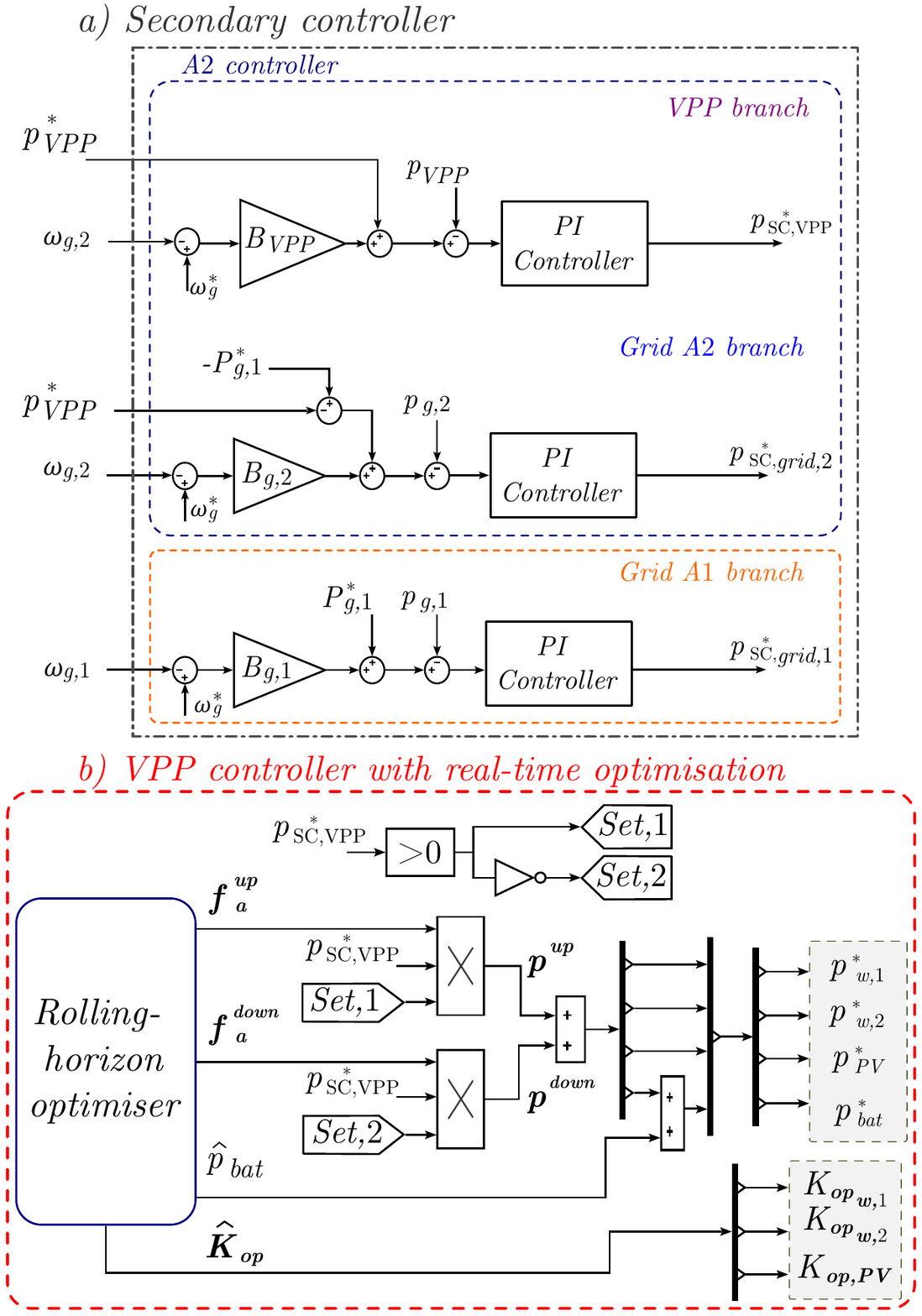}
\vspace{-0.1cm}
\caption{Controller block diagram: a) secondary controller; b) VPP  with real-time optimisation. }
\vspace{-0.4cm}
\label{fig:Secondary_Controller}
\end{figure}

Fig.~\ref{fig:Secondary_Controller}(b) shows a detailed block diagram of the proposed VPP controller.
Its main role is to calculate the power references and reserve allocation for the VPP elements, taking into consideration the availability of the resources in order to maximise the efficient operation of the VPP.

The rolling-horizon optimisation dynamically computes two sets of variables: the optimal power allocation among the VPP units and the availability factors for the provision of frequency restoration services. 
The first set defines the operating condition for each unit, given by the variables $\bm{\hat{K}_{op}}=\left[K_{op,{w,1}}\;K_{op,{w,2}}\;K_{op,{PV}}\right]$ and $\hat{p}_{bat}$: 
$\bm{\hat{K}_{op}}\in[0,1]$ determines the operating factor for wind and PV, i.e., the fraction of the maximum available power that each source injects; and $\hat{p}_{bat}$ determines the power output of the battery.
To compute these values, the optimiser considers forecasts of the load demand and the key parameters governing power generation and storage (i.e., irradiance, wind speed, and SOC).
In addition, the optimiser incorporates relationships that account for degradation costs, operating costs, and high penalties for improper unit operation, e.g., deep charge/discharge cycles.

The second set governs the provision of frequency restoration service.
In order to achieve this, the optimiser defines a set of availability factors, $\bm{f_a}=[f_{a_{w,1}}\;f_{a_{w,2}}\;f_{a_{PV}}\;f_{a_{bat}}]$, differentiating between those for the provision of upward reserve $(\bm{f^{up}_a})$ and downward reserve $(\bm{f^{down}_a})$. 
These factors depend on the reserve capacity of each source at the operating point.
Depending on whether the VPP is providing upward or downward reserve, the VPP controller has two operating modes. 
A binary signal ($Set$) is used to distinguish between these two cases ($Set=0$ for upward reserve, and $Set=1$ for downward reserve). 
The power requested from the VPP for frequency restoration $p_{SC,VPP}^*$ is shared between the VPP units according to such availability factors.
\section{Power System Modelling}
\label{sec.power.sys.model}
This section introduces the fundamentals of the models used to describe the system in Fig.~\ref{fig:VPP_Single_Diagram}. 
They are presented here for the sake of completeness (see \cite{Marco_IECON_2025,AccessModels} for details).
\subsubsection{Equivalent Model of Areas}
Each electric network area is modelled as a first-order swing equation, with a speed controller (droop) and loads, which are represented as an input~\cite{P_Systems_Dynamics}.
This input will be later used to model load profiles. 
\subsubsection{Wind Power Generators}
A shaft with a generator and an equivalent inertia ($H$) is used to describe the wind farms.
The output is electrically coupled to the grid using a power source~\cite{Marco_IECON_2025}.
It includes a speed controller that calculates the operating point taking into consideration the wind speed and the power reference ($p_{w}^*$).
The latter is determined by an operating power point tracker (OPPT).
The power coefficient $c_p$ is estimated using the mathematical approximation introduced in~\cite{Cp_calculate}.
The grid interface is done with a standard grid-following converter that includes an \textit{LCL} filter to eliminate high-frequency harmonics.
Its control system is based on a $dq$ current controller, active and reactive power controllers, and dc-voltage regulator.
\subsubsection{PV Generator}
A controlled power source is used to model PV generators, where the power output is calculated as a function of the irradiance~\cite{PV_model1}. 
The PV generators are interfaced to the grid via a dc-dc and a dc-ac converter. 
The grid interface includes an additional \textit{LCL} filter.
For calculating the PV operating point, an OPPT is used. 
This algorithm considers the irradiance and the power reference ($p_{PV}^*$) sent from the VPP set-point and the frequency restoration control.   
\subsubsection{Battery Energy Storage System}
The battery is modelled using a series resistance and an $RC$ branch, where the open-circuit voltage depends on the SOC~\cite{Batteries_mod2}.
The BESS is interfaced with the grid using a dc-ac voltage source converter, equipped with an \textit{LCL}
filter~\cite{VSC_book}. 
A grid-forming control is included in the control system to support the grid.
This controller is formed by a virtual impedance loop, active and reactive power controllers, and standard $dq$ current and voltage regulators~\cite{Z_virtual_Jav}.
It follows the reference $p^*_{bat}$, resulting from the VPP setpoint and the frequency restoration control. 
\subsubsection{Transformers, Lines, and Loads}
Transformers are represented with series equivalent circuits formed by an inductor and a resistor. 
The apparent power consumed by loads is represented with equivalent resistor-inductor circuits, in parallel connection.
For representing lines, a gamma model is used. 
\section{Rolling-Horizon Optimiser}
\label{sec.proposed.optimiser}
\subsection{Optimisation Framework}
This work presents the implementation of a rolling-horizon optimiser for the control of a VPP that provides frequency restoration support in a multi-area power system.
The optimiser continuously determines the optimal power sharing among the VPP units in order to follow the power and reserve references scheduled in a day-ahead market (DAM).
The optimisation process aims to maximize the utilization of the available energy resources while considering operational and degradation constraints.
As shown in Fig.~\ref{fig:VPP_Single_Diagram}, the inputs to the rolling-horizon optimiser are the wind speed, solar irradiance, SOC, the VPP power reference $(p^*_{VPP})$, and the committed reserves ($R^{up^*}_{VPP}$,~$R^{down^*}_{VPP}$).
The latter are defined by a day-ahead optimisation, briefly described next.

\subsection{Day-Ahead Optimisation}
This stage aims at maximising the economic profits of the VPP from trading in the energy and secondary reserve markets.
The objective function can be defined as:
\begin{equation}
\mathrm{max}\sum _k
{\pi_{1,k} 
\cdot 
{p}_{VPP,k}^{*} +
\pi_{2,k}^{up,down} 
\cdot 
{R}_{VPP,k}^{up,down^*}},
\label{eq.ObjFuncDAM}
\end{equation}
where $\pi_{1,k}$ and $\pi_{2,k}^{up,down}$ are the energy and upward and downward reserve prices at the $k^{\textrm{th}}$ time step.
This stage optimises the VPP operation for a 1-day horizon with a time step of 1 minute.
To comply with market requirements, both $p_{VPP}^*$ and $R_{VPP}^{up,down^*}$ are forced to be constant within each market period, i.e., each quarter-hour~\cite{BOE_137_250424}.

This optimisation stage considers the forecast values of load, wind speed and irradiance.
It includes typical constraints such as power balance and power and energy constraints, similar to the rolling-horizon optimiser that
will be shown next. 
Since this stage is not the main focus of this work and its purpose is only to obtain realistic power dispatch and reserve commitment profiles, more details can be found in~\cite{Diego_TSTE_Central}, extended here to include the wind farms. 
\subsection{Rolling-Horizon Optimisation}
In the following, variables marked with a tilde (~$\tilde{\cdot}$~) represent forecasted values, and those marked with a hat (~$\hat{\cdot}$~) represent decision variables of the rolling-horizon optimiser.
Among the latter, the main outputs of the optimiser are the VPP active power $\hat{p}_{VPP}$, the upward and downward reserves $\hat{R}^{up,down}_{VPP}$, and the operating factors $\bm{\hat{K}_{op}}$ of the renewable units.
\subsubsection{Objective Function}
\begin{figure}[!b]
\centering
\vspace{-0.2cm}
\includegraphics[width=1\columnwidth]{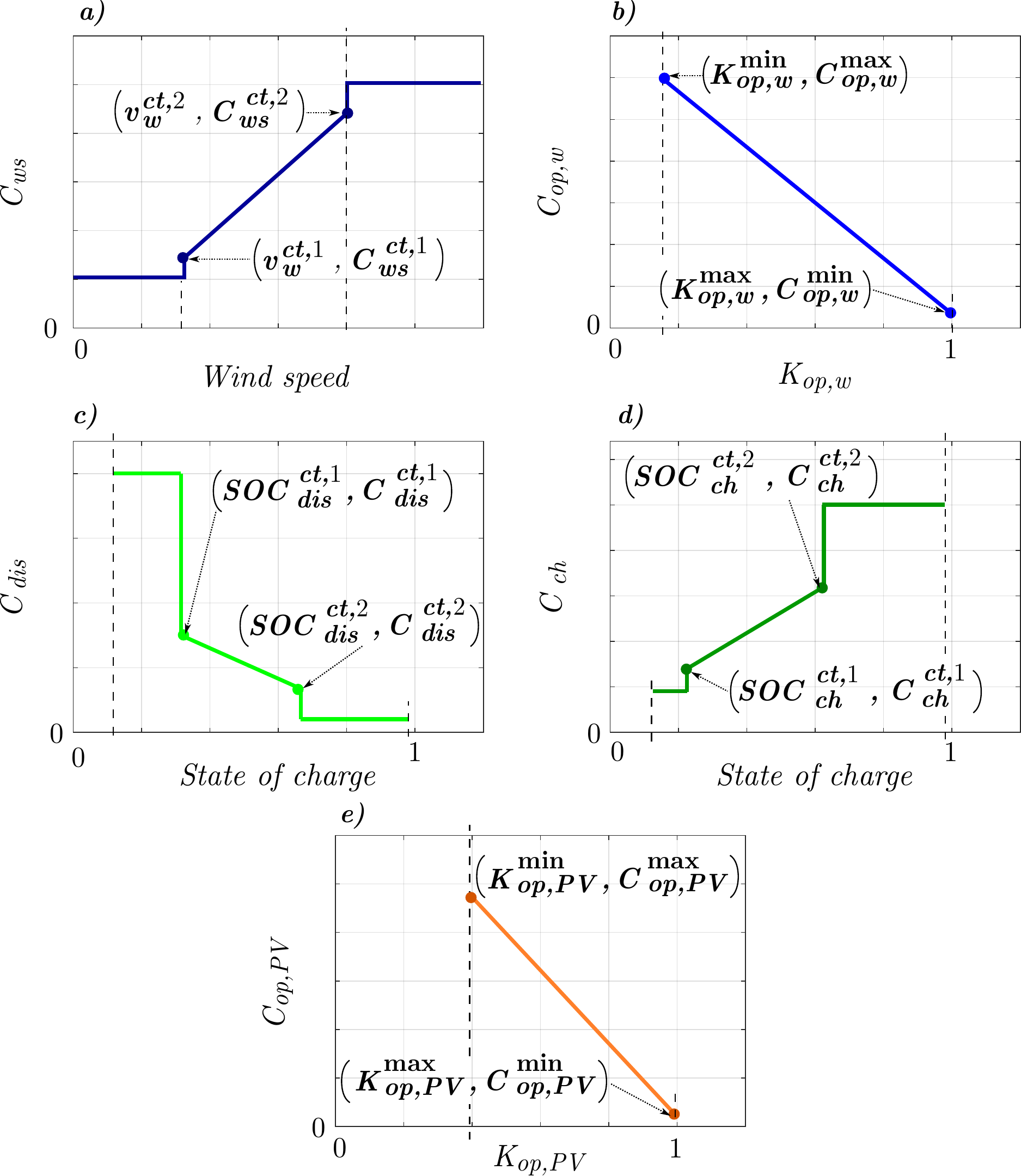}
\caption{Cost-function curves used to model the VPP units: (a) $C_{ws}$ and (b) $C_{op,w}$ for the wind farms; (c) $C_{dis}$ and (d) $C_{ch}$ for the BESS; and (e) $C_{op,PV}$ for the PV plant.}
\label{fig:Cost_Curves}
\end{figure}

The objective function is defined as the sum of all cost terms. 
\begin{equation}
\begin{aligned}
\mathrm{min}\sum_{k} 
\Big[ 
&\sum_{i=1}^{2}(C_{op,w,i} + C_{ws,i}) + C_{dis} + C_{ch} +C_{op,PV}\\&+
C_{\Delta p} + C_{\Delta {R}}^{up,down}\Big], 
\end{aligned}
\label{eq:OF_rolling_opt}
\end{equation}
where $C_{op,w,i}$ and $C_{ws,i}$ are the costs related to the operating point of the wind turbines and wind speed, respectively, for the $i^{th}$ wind farm; $C_{dis,ch}$ are the costs associated with the degradation of the BESS when it discharges and charges; $C_{op,PV}$ is the cost associated with the operating point of the PV plant; and $C_{\Delta p}$ and $C_{\Delta {R}}^{up,down}$ are penalisations associated with the error tracking of the power and reserve references.
The summation in~(\ref{eq:OF_rolling_opt}) extends over all time steps $k$ of the rolling optimisation horizon.
For simplicity, the time-step index $k$ is omitted and is shown only where temporal coupling is relevant (e.g., SOC computation).
Fig.~\ref{fig:Cost_Curves} illustrates the shape of the cost terms and the parameters that define them. 
The details are illustrated next.

\paragraph{Costs of Wind Generation}
For simplicity, the wind farm index $i$ is omitted here, as the model is identical for both.
Operation under high wind-speed conditions is one of the primary sources of wind turbine degradation.
To account for this effect, a term that depends on the wind speed is introduced.
Fig.~\ref{fig:Cost_Curves}(a) shows the curve used to model this cost ($C_{ws}$), which can be mathematically expressed as follows: 
\begin{align}
C_{ws} 
=
\begin{cases}
C^{\;\min}_{ws}, & v_{w} \in I_{1}, \\
m_{ws} (v_{w}-v^{ct,1}_{w} ) 
+ 
C^{\;ct,1}_{ws}, & v_{w} \in I_{2}, \\
C^{\;\max}_{ws}, & v_{w} \in I_3, \\
\end{cases}
\end{align}
where the slope is:
\begin{align}
m_{w{s}} 
&=
({C^{{\;ct,2}}_{w{s}} - 
C^{\;ct,1}_{ws}})
/ 
(v^{ct,2}_{w} - v^{ct,1}_{w}), 
\end{align}
with $v^{ct,2}_{w}>v^{ct,1}_{w}$, $C^{{\;ct,2}}_{w{s}}> 
C^{\;ct,1}_{ws}$.
In these equations $v_{w}$ is the wind speed, and $v^{ct,1}_{w}$ and $v^{ct,2}_{w}$ define the thresholds that define the intervals $I_1$, $I_2$, and $I_3$, corresponding to $v_{w}$~$<$~$v^{ct,1}_{w}$, $v^{ct,1}_{w} \leq v_{w} \leq v^{ct,2}_{w} $, and $v_{w}>v^{ct,2}_{w}$, respectively; 
$C^{{\;ct,1}}_{ws}$ and $C^{{\;ct,2}}_{ws}$ denote the costs associated with wind speeds $v^{ct,1}_{w}$ and $v^{ct,2}_{w}$, respectively.

Since one of the objectives of the optimisation is to maximise the utilisation of the available resources, a penalisation cost associated with renewable generation curtailment is included.
As shown in Fig.~\ref{fig:Cost_Curves}(b), this operating cost ($C_{{op},{w}}$) decreases as ${K}_{op,w}$ increases, and is calculated as follows:
\begin{align}
C_{{op},{w}} 
&= 
m_{{op},{w}} \cdot  
({K}_{{op},{w}} - 
K^{\;\min}_{{op},{w}}) + 
C^{\;\max}_{{op},{w}}, \\
m_{{op},{w}} 
&= 
\frac{C^{{\;\min}}_{{op},{w}} - 
C^{{\;\max}}_{{op},{w}}}
{K^{\;\max}_{{op},{w}} - 
K^{\;\min}_{{op},{w}}}, 
\end{align}
where $C^{{\;\min}}_{{op},{w}}$ and $C^{\;\max}_{{op},{w}}$ are the minimum and maximum values of $C_{{op},{w}}$, respectively. 
Similarly $K^{\;\min}_{{op},{w}}$ and $K^{\;\max}_{{op},{w}}$ are the minimum and maximum values of $K_{{op},{w}}$, respectively.

\paragraph{Costs of BESS Units}
%
%The main cost associated with BESS units are %due to their use.
The most significant BESS costs are associated with operation and degradation. 
These costs are strongly influenced by both the operating mode (charging or discharging) and the SOC range in which the batteries operate.

During battery discharging, the cost term $C_{dis}$ is modelled as a piecewise linear function of the SOC. 
In general, this cost decreases as the SOC increases. 
Also, high penalties are imposed at low SOC levels to prevent deep discharges.
The curve that models this behaviour is presented in Fig.~\ref{fig:Cost_Curves}(c), and can be mathematically expressed as follows: 
\begin{equation}
\label{Cost_dis}
\resizebox{0.45\textwidth}{!}{$
C_{dis} = 
\begin{cases}
C^{\;\max}_{dis}, 
& SOC \in I_{1,dis}, \\
m_{dis}
(SOC-SOC^{\;ct,1}_{dis}) +  C^{\;ct,1}_{dis}, 
& SOC \in I_{2,dis}, \\
C^{\;\min}_{dis}, 
& SOC \in I_{3,dis},
\end{cases}
$}
\end{equation}
where the slope is given by: 
\begin{equation}
\label{Cost_dis_slope}
m_{dis} 
=
\frac
{{C^{\;ct,2}_{dis}-C^{\;ct,1}_{dis}}}
{{SOC^{\;ct,2}_{dis}-SOC^{\;ct,1}_{dis}}},
\end{equation}
with $SOC^{ \;ct,1}_{dis}<SOC^{\;ct,2}_{dis}$ and $C^{\;ct,1}_{dis}>C^{\;ct,2}_{dis}$.
The variables $C^{\;\max}_{dis}$ and $C^{\;\min}_{dis}$ denote the maximum and minimum values of $C_{dis}$, respectively;
the parameters $SOC^{\;ct,1}_{dis}$ and $SOC^{\;ct,2}_{dis}$ define the boundaries of the intervals $I_{1,dis}$, $I_{2,dis}$, and $I_{3,dis}$, corresponding to  $SOC<SOC^{\;ct,1}_{dis}$, $SOC^{\;ct,1}_{dis} \leq SOC \leq SOC^{\;ct,2}_{dis}$, and  $SOC > SOC^{\;ct,2}_{dis} $, respectively.

Accordingly, during battery charging, the cost $C_{ch}$ introduces higher penalties at high SOC levels to prevent overcharging.
These relationships are illustrated in Fig.~\ref{fig:Cost_Curves}(d) and are represented as follows:
\begin{align}
C_{ch} =
\begin{cases}
C^{\;\min}_{ch}, & SOC \in I_{1,ch}, \\
m_{ch}
(SOC-SOC^{ct,1}_{ch}) + C^{ct,1}_{ch}, & SOC \in I_{2,ch}, \\
C^{\;\max}_{ch}, & SOC \in I_{3,ch},
\end{cases}
\end{align}
where the slope is given by:
\begin{align}
m_{ch} 
=
\frac{{C^{\;ct,2}_{ch}-C^{\;ct,1}_{ch}}}
{{SOC^{\;ct,2}_{ch}-SOC^{\;ct,1}_{ch}}},
\end{align}
and $SOC^{\;ct,1}_{ch}<SOC^{ct,2}_{ch}$ and $C^{\;ct,1}_{ch}<C^{\;ct,2}_{ch}$.
The notation used here is equivalent to that in (\ref{Cost_dis}) and (\ref{Cost_dis_slope}), except for the subscript $ch$, which refers to ``charging mode''.

\paragraph{Costs of PV Units}
The operating cost relationship for the PV generation is illustrated in \ref{fig:Cost_Curves}(e). 
Similarly to wind generation operating costs, $(C_{{op,PV}})$ is modelled as follows:
\begin{align}
C_{{op,PV}} 
&= 
m_{{op},{PV}} 
\left ({K}_{{op,PV}}-K^{\;\min}_{{op,PV}} \right) + C^{\;\max}_{{op,PV}}, \\
m_{{op},{PV}} 
&= 
\frac{C^{{\;\min}}_{{op},{PV}} - C^{{\;\max}}_{{op},{PV}}}
{K^{\;\max}_{{op},{PV}} - K^{\;\min}_{{op},{PV}}},
\end{align}
where $C^{{\;\min}}_{{op,PV}}$ and $C^{{\;\max}}_{{op},{PV}}$ are the minimum and maximum values of $C_{{op,PV}}$, respectively; 
$K^{\;\min}_{{op,PV}}$ and $K^{\;\max}_{{op,PV}}$ are the minimum and maximum values of $K_{{op,PV}}$, respectively.
Since the degradation costs associated with PV generation are negligible, they are not considered in this study.

\paragraph{Penalty on Active Power Reference Tracking Error} 
To ensure that $\hat{p}_{VPP}$ closely tracks $p^*_{VPP}$, a penalty term ($C_{\Delta{p}}$) is added to the objective function.
It is modelled as follows:
\begin{align}
\label{eq.power_penalty1}
C_{\Delta{p}}
&=
a_1 \cdot \pi_1 \cdot \Delta \hat{p}_{VPP} ,\\
\label{eq.power_penalty2}
\Delta \hat{p}_{VPP} &= |p^*_{VPP}-\hat{p}_{VPP}|,
\end{align}
where $a_1$ is a penalty factor and $\Delta \hat{p}_{VPP}$ is the active power tracking error.
\paragraph{Penalty for Reserve Reference Tracking Error}
Similar to the case above, to ensure that $\hat{R}_{VPP}^{up,down}$ closely tracks $R^{*\;up,down}_{VPP}$, a penalty term ($C_{\Delta{R}}^{up,down}$) is added to the objective function.
It is modelled as follows:
\begin{align}
C^{up,down}_{\Delta R}
&=
a_2 \cdot \Delta \hat{R}^{up,down}_{VPP}\\
\label{eq.reserve_penalty}
\Delta \hat{R}_{VPP}^{up,down} &= \max(0,R^{*\;up,down}_{VPP}-\hat{R}^{up,down}_{VPP}),
\end{align}
where $a_2$ is a penalty factor, and $\Delta \hat{R}_{VPP}^{up,down}$ are the upward/downward reserve tracking errors.
It is noted that this penalty applies only when the actual reserve ($\hat{R}^{up,down}_{VPP}$) is less than the scheduled one ($R^{*\;up,down}_{VPP}$). 

\subsubsection{Constraints}
\paragraph{Modelling of Resources}
The maximum power generation forecast for each renewable source in the VPP is:
\begin{align}
\label{P_max_wind}
\tilde{p}^{\;\max}_{w,i} 
&= 
\frac{1}{2} n_{w,i} \; \rho\; c^{\max}_p A_p \; \tilde{v}_{w,i}^3, \\
\label{P_max_PV}
\tilde{p}^{\;\max}_{PV} 
&= 
\frac{\tilde{G}}{1000}P^{\;nom}_{PV},
\end{align}
and the maximum available power from the battery, both for charging and discharging, is:
\begin{align}
\label{P_max_bat}
{P}^{\;\max}_{bat} 
&= 
P^{\;nom}_{bat}.
\end{align}

In~(\ref{P_max_wind}), $n_{w,i}$ denotes the number of wind turbines in the $i^{\mathrm{th}}$ wind farm, $\rho$ is the air density, $c^{\max}_p$ is the maximum power coefficient, $A_p$ is the area covered by the blades of a wind turbine, and $\tilde{v}_{w,i}$ is the wind speed forecast at location $i$.
It is noted that the wind speed, and hence the corresponding forecast, differs for different wind farm locations.
In~(\ref{P_max_PV}), $\tilde{G}$ denotes the solar irradiance forecast and $P^{\;nom}_{PV}$ is the rated power of the PV generator. 
Finally, in (\ref{P_max_bat}), $P^{\;nom}_{bat}$ is the BESS rated power.

\paragraph{Operation-Point Constraints}
For renewable generators, their active power output is computed from the variable $\bm{\hat{K}_{op}}=\left[K_{op,{w,1}}\;K_{op,{w,2}}\;K_{op,{PV}}\right]$.
This can be mathematically expressed as follows: 
\begin{align}
\hat{p}_{w,i} 
&= 
{K}_{op,{w,i}} \cdot \tilde{p}^{\;\max}_{w,i}, \\
\hat{p}_{PV} 
&= 
{K}_{op,{PV}} \cdot \tilde{p}^{\;\max}_{PV}.
\end{align}
The values of $\bm{\hat{K}_{op}}$ elements must always remain between a minimum ($K^{\;\min}_{op,{w,i}}$, $K^{\;\min}_{op,{PV}}$) and one.
In addition, a constraint is introduced to ensure that the active power operating point is not above the rated power of the wind farms when wind speed is high:
\begin{align}
\label{K_op_ranges}
{\rm if} \;\tilde{p}^{\;\max}_{w,i} 
> 
P^{\;nom}_{w,i}, {\; \rm then\;} 
K_{op,{w,i}} 
\leq 
\frac{P^{\;nom}_{w,i}}{\tilde{p}^{\;\max}_{w,i}}.
\end{align}

\paragraph{Active Power Balance}
The VPP power balance constraint is expressed as:
\begin{align}
\label{Power_Balance}
\hat{p}_{VPP} 
&= 
\hat{p}_{w,1} + \hat{p}_{w,2} 
+ 
\hat{p}_{PV} + \hat{p}_{bat} 
- \tilde{p}_{load},
\end{align}
where $\hat{p}_{VPP}$, $\hat{p}_{w,1}$, $\hat{p}_{w,2}$, $\hat{p}_{PV}$, and $\hat{p}_{bat}$ are the power injected by the VPP, wind generators, PV, and BESS, respectively.
The BESS follows the generator sign convention, i.e., $p_{bat}$ is positive when discharging and negative when charging.
To track the power reference scheduled in the DAM, power deviations between $\hat{p}_{VPP}$ and $p^*_{VPP}$ are penalised as in~(\ref{eq.power_penalty1})-(\ref{eq.power_penalty2}).
\paragraph{Reserve Availability}
The upward reserve of each $j^{\mathrm{th}}$ renewable unit $(\hat{R}^{up}_{j})$ and of the BESS is computed taking into consideration their maximum active power outputs:
\begin{align}
\label{R_ups}
\hat{R}^{up}_{j} 
&= 
\tilde{p}^{\;\max}_{j} 
- 
\hat{p}_{j}, \;\;\;\;\;\; j \in \{w,1;w,2;PV\},\\
\hat{R}^{up}_{bat} 
&= 
P^{\max}_{bat} 
- 
\hat{p}_{bat}.
\end{align}
To compute the downward reserve, the minimum operating factors $K^{\min}_{op}$ are considered for wind and PV power plants.
The available reserve of each renewable source ($\hat{R}^{down}_{j}$) and of the BESS is then calculated as follows:
\begin{align}
\label{R_downs_1}
\hat{R}^{down}_{j} 
&= 
\; \hat{p}_{j} - \left(K^{\min}_{op,j}\;
\tilde{p}^{\;\max}_{j}\right), \;\;\; j \in \{w,1;w,2;PV\},\\
\hat{R}^{down}_{bat} 
&= 
\; \hat{p}_{bat} 
+ 
P^{\;\max}_{bat}.
\end{align}
Then, the total upward and downward reserves are:
\begin{align}
\label{R_u}
\hat{R}^{up}_{VPP} 
&= 
\hat{R}^{up}_{w,1} + 
\hat{R}^{up}_{w,2} + 
\hat{R}^{up}_{PV} + 
\hat{R}^{up}_{bat}, \\ 
\label{R_d}
\hat{R}^{down}_{VPP} 
&= 
\hat{R}^{down}_{w,1} + 
\hat{R}^{down}_{w,2}  + 
\hat{R}^{down}_{PV} + 
\hat{R}^{down}_{bat}.
\end{align}

To comply with the reserve commitments, the upward and downward reserve capacities of the VPP should be greater than or equal to the reference.
Otherwise, deviations are calculated as in~(\ref{eq.reserve_penalty}) and penalised.

Once the optimal reserve values are calculated in the optimisation problem, the upward and downward reserve availability factors of each VPP unit are calculated as follows:
\begin{align}
f^{up}_{a,j} 
&= 
\hat{R}^{up}_{j} \; / \hat{R}^{up}_{VPP} \;\;\;\;\;\;  j \in \{w,1;w,2;PV;bat\}, \\
f^{down}_{a,j} 
&= 
\hat{R}^{down}_{j} / \hat{R}^{down}_{VPP}.
\end{align}
\paragraph{BESS Operation}
The BESS active power is divided into charging ($\hat{p}_{{b,ch}}$) and discharging ($\hat{p}_{{b,dis}}$) powers:
\begin{align}
\label{p_bat}
\hat{p}_{{bat}} & = 
\hat{p}_{{b,dis}} - \hat{p}_{{b,ch}}. 
\end{align}
Both powers are non-negative,  and a complementarity constraint avoids simultaneous charging and discharging.

The SOC is estimated considering the charging and discharging modes, and applying the following relationship:
\begin{equation}
SOC(k+1)
=
SOC(k) 
+ 
\frac{\eta \; \hat{p}_{b,ch}(k)}{Q_n}\Delta t -
\frac{\hat{p}_{b,dis}(k)}{\eta \; Q_n}\Delta t,
\end{equation}
where $\eta$ is the battery efficiency (assumed constant and equal for charging and discharging), $Q_n$ denotes the BESS nominal capacity, and $\Delta t$ corresponds to the sampling period of the optimisation problem. 
To ensure the operation of the BESS within its limits, the SOC values must be constrained:
\begin{equation}
SOC^{\;\min} 
\leq SOC 
\leq SOC^{\;\max}.
\end{equation}
\section{Results}
\label{sec.results}
This section is focused on presenting the simulation results of the VPP to validate the feasibility of the proposed VPP controller.
To this end, the two-area power system depicted in Fig.~\ref{fig:VPP_Single_Diagram} is used.
The system is simulated in MATLAB/Simulink, and the minimisation problem is defined using YALMIP~\cite{Yalmip} and solved with Gurobi.
The system parameters are shown in Table~\ref{tab:Parameters}.
It is worth noting that two minimum values are considered for $K^\mathrm{min}_{op,w}$: $K^\mathrm{min}_{op,w}=0.5$ as the standard value, and $K^{\min}_{op,w}=0.15$, applied only when the estimated maximum wind turbine power exceeds its rated power.
The DAM optimises the VPP operation in a 1-day horizon, with a time step of 1 minute. 
The rolling horizon optimisation employs the same time step over a 10-minute horizon.
\begin{table}[!t]
\vspace{-0.2cm}
\caption{Model and Optimisation Parameters}
\vspace{-0.1cm}
\centering
\resizebox{\columnwidth}{!}{
\renewcommand{\arraystretch}{1.4}
\begin{tabular}{lc|lc}\toprule
\multicolumn{2}{c|}{\textbf{Wind farms}} & \multicolumn{2}{c}{\textbf{BESS}} \\\hline
 $\eta_{w,1}$  & 6  & $Q_n$ (MWh) & 1382 \\
 $\eta_{w,2}$ &  8 & $SOC_{\max/\min}$ & 0.95/0.2 \\
$c^{\max}_p$ & 0.47 & $\eta$ & 0.95 \\
$K^{\min}_{op,w}$ & 0.5\;/\;0.15 & $SOC^{ct,1}_{dis}$, $SOC^{ct,2}_{dis}$ & 0.4, 0.75 \\
$v^{ct,1}_{w}$, $v^{ct,2}_{w}$ (m/s) & 15, 20 & $SOC^{ct,1}_{ch}$, $SOC^{ct,2}_{ch}$ & 0.2, 0.7 \\
$C^{\min}_{ws}$, $C^{\max}_{ws}$ (€) & 0.5, 2.5 & $C^{ct,1}_{dis}$, $C^{ct,2}_{dis}$ (€) & 1.5, 0.68 \\
$C^{\min}_{op,w}$, $C^{\max}_{op,w}$ (€) & 0.1, 3 & $C^{ct,1}_{ch}$, $C^{ct,2}_{ch}$ (€) & 0.81, 1.6 \\[2pt]\hline\hline
\multicolumn{2}{c|}{\textbf{PV plant}} & \multicolumn{2}{c}{\textbf{General}} \\\hline
$K^{\min}_{op,PV}$ & 0.7 & $P^{nom}_{\{w1,w2,PV,bat\}}$ (MW) & 60 \\
$C^{\min}_{op,PV}$, $C^{\max}_{op,PV}$ (€) & 0.1, 2.5 &  $a_1$, $a_2$ & 1.2, 10 \\
\bottomrule
\end{tabular}
}
\label{tab:Parameters}
\vspace{-0.3cm}
\end{table}

Two case studies are presented and analysed. 
In the first one, the day-ahead forecasts (i.e., wind speed, irradiance, load profile) coincide with the real-time operating values. 
In contrast, in the second case, deviations between the forecasted and actual values are considered.
In other words, in the first case, the evaluation of frequency support and reserve allocation is performed under perfect forecast conditions, while in the second case the evaluation is performed under more realistic operation conditions.

Fig.~\ref{fig:Powers_DAM} shows the forecasted power profiles and the active power references calculated by the DAM optimisation. 
It can be seen that there are two large variations (at $t=9$~h and $t=18$~h, approximately), which correspond to sunrise and sunset.
The other large variations are less pronounced (at $t=2$~h and $t=6$~h, approximately), and correspond to changes in the solution, mainly because the BESS changes its operating conditions. 
These values will be used as references in the following case studies. 
\subsection{Case 1: Operation Under Ideal Forecast Conditions}
\label{Case_1_Results}
\begin{figure}[!t]
\centering
\vspace{-0.1cm}
\includegraphics[width=0.97\columnwidth]{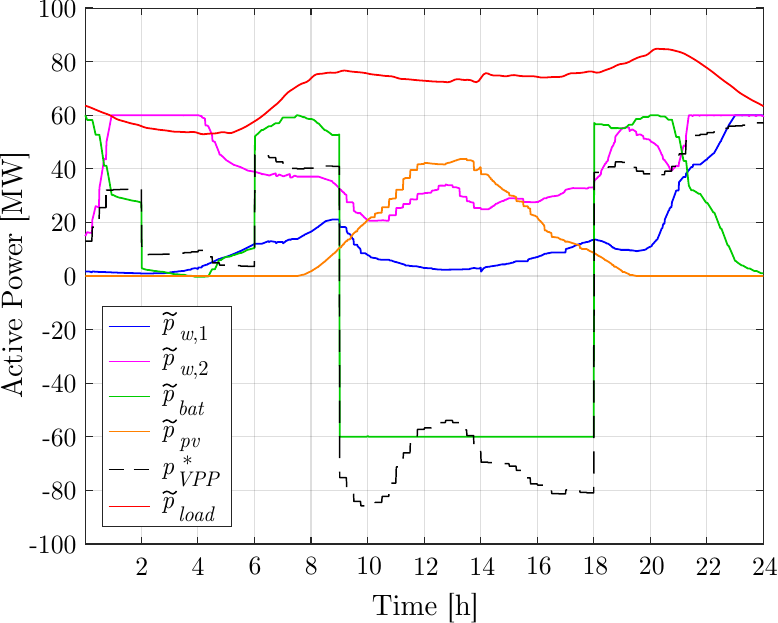}
\vspace{-0.3cm}
\caption{Active power profiles resulting from the DAM optimisation.}
\vspace{-0.3cm}
\label{fig:Powers_DAM}
\end{figure}

\begin{figure}[!t]
\centering
\vspace{-0.2cm}
\includegraphics[width=0.97\columnwidth]{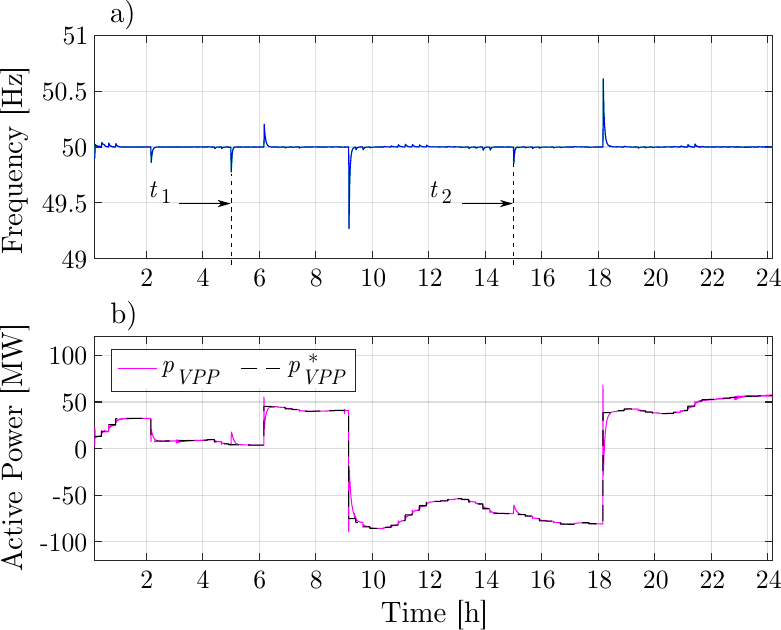}
\vspace{-0.2cm}
\caption{Time response of the power system under perfect forecast conditions: a)~frequency; b)~active power injected by the VPP.}
\vspace{-0.3cm}
\label{fig:Freq_and_P_vpp_case1_V0}
\end{figure}

\begin{figure}[!t]
\centering
\vspace{-0.1cm}
\includegraphics[width=0.99\columnwidth]{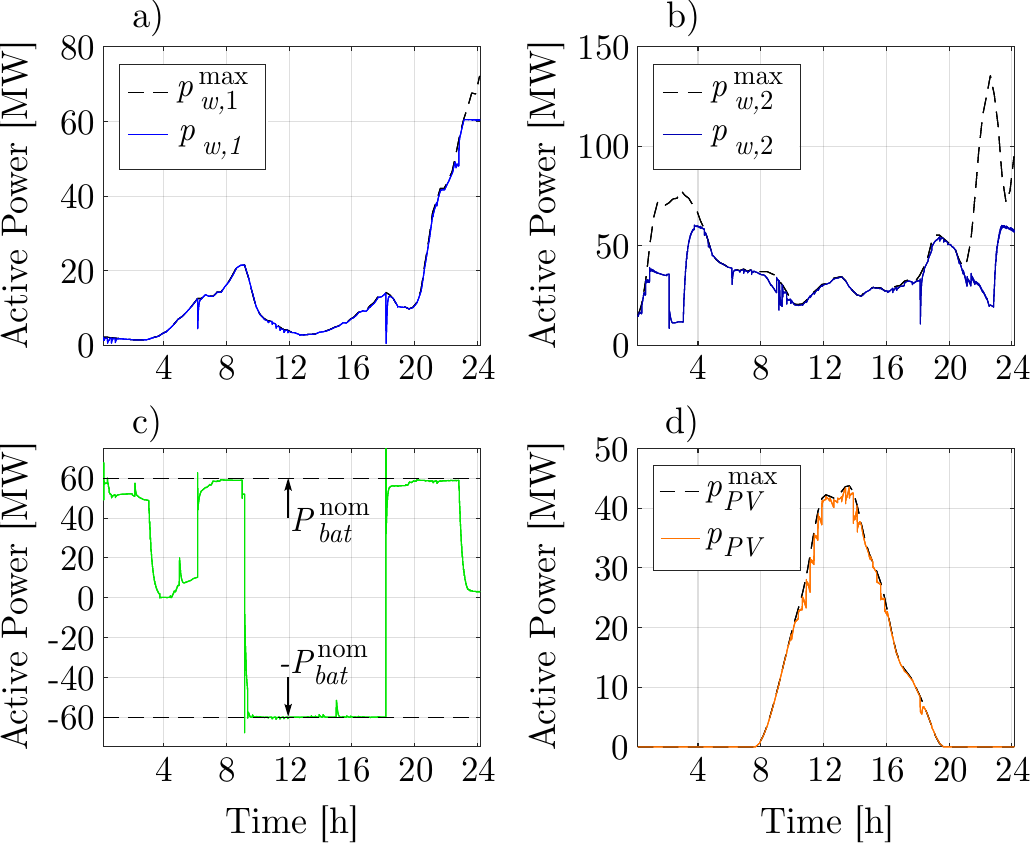}
\vspace{-0.4cm}
\caption{Active power profiles injected by the single VPP units: a) wind farm~1; b) wind farm~2; c) PV plant; c) BESS.}
\vspace{-0.1cm}
\label{fig:Powers_VPP_Units_case1}
\end{figure}

\begin{figure}[!t]
\centering
\vspace{-0.1cm}
\includegraphics[width=0.97\columnwidth]{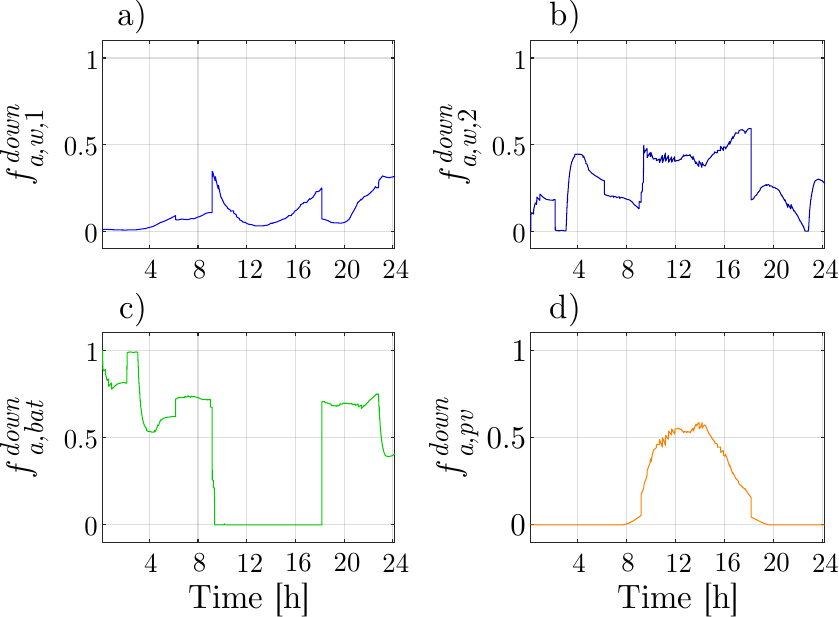}
\vspace{-0.1cm}
\caption{Downward availability factors of: a) wind farm 1; b) wind farm 2; c) PV plant; d) BESS.}
\vspace{-0.3cm}
\label{fig:fa_down}
\end{figure}

\begin{figure}[!t]
\centering
\vspace{-0.2cm}
\includegraphics[width=0.97\columnwidth]{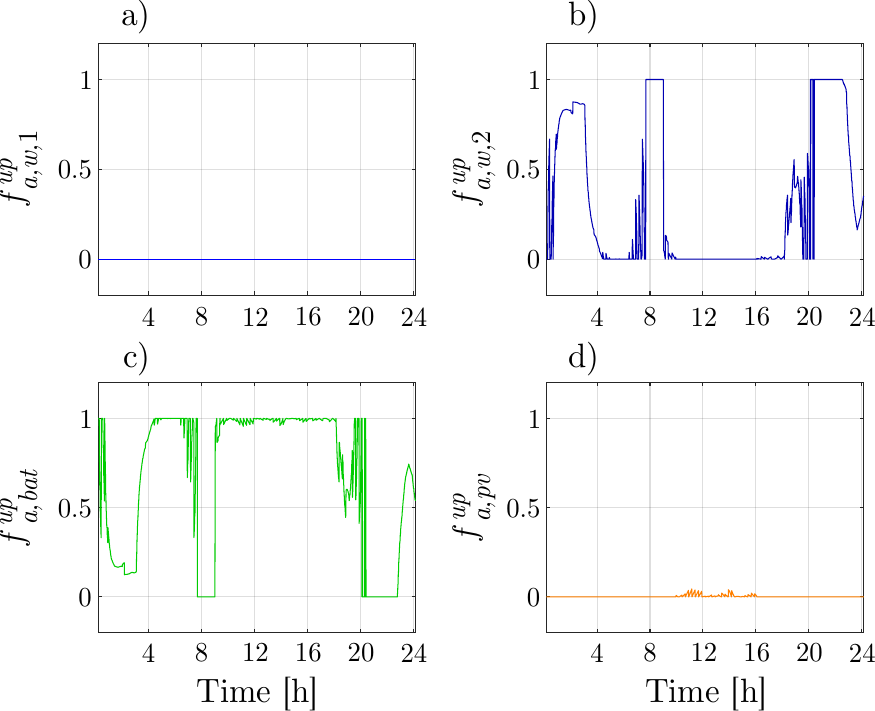}
\vspace{-0.2cm}
\caption{Upward availability factors of: a) wind farm 1; b) wind farm 2; c) PV plant; d) BESS.}
\vspace{-0.2cm}
\label{fig:fa_up}
\end{figure}

\paragraph{Frequency and VPP Active Power Response}
The network frequency and the active power delivered by the VPP are shown in Fig.~\ref{fig:Freq_and_P_vpp_case1_V0}(a) and Fig.~\ref{fig:Freq_and_P_vpp_case1_V0}(b), respectively. 
The frequency deviates from the nominal value when the VPP active power is modified.
Step-changes of $0.2$~pu are applied in the load of grid 1 (at $t_1$ = 5~h) and 2 (at $t_2$ = 15~h).
It can be observed that, in both cases, frequency is quickly restored to its reference value. 
The largest frequency variations occur at $t=9$~h and $t=18$~h, as this instant coincides with the largest variations in the reference values calculated in the DAM optimisation.
In addition, the active power injected by the VPP follows its reference, and it rapidly compensates for the system load changes at $t_1$ and $t_2$.
These results validate the main functionalities of the proposed controller. 
\paragraph{Generation and Storage of VPP Units}
Fig. ~\ref{fig:Powers_VPP_Units_case1} shows the active power injected by the generators and storage units of the VPP, together with their corresponding power limits.
Most of the time, the signals exhibit smooth variations. 
However, significant changes can be observed at specific time instants, which are associated with the largest variations in the active-power references determined in the DAM optimisation.
These variations can be smoothed using ramps, although smoothing should not be excessive to ensure that the optimal solution is not modified.
Regarding wind generation, Fig.~\ref{fig:Powers_VPP_Units_case1}(a) shows that the wind farm at node 1 operates almost always at its maximum capacity throughout the considered period, except when it saturates at $t=23$~h.
In contrast, Fig.~\ref{fig:Powers_VPP_Units_case1}(b) shows that the active power generated by wind farm at node 2 deviates more from its theoretical maximum.
This happens because the wind speed at node 2 is generally higher than at node 1 (e.g., see the saturation between 1~h and 4~h in Fig.~\ref{fig:Powers_DAM} and the wind speed in Fig.~\ref{fig:Resources}(b)) and the allocation performed by the rolling-horizon optimiser.
Fig.~\ref{fig:Powers_VPP_Units_case1}(c) shows that the BESS operation remains close to the profile scheduled in the DAM.
The transients resulting from the reference changes in the DAM, as well as those from the system load changes, can also be observed.
Finally, Fig.~\ref{fig:Powers_VPP_Units_case1}(d) shows that the PV plant operates close to its maximum available power throughout the daytime period.
\paragraph{Downward and Upward Availability Factors}
Fig.~\ref{fig:fa_down} and Fig.~\ref{fig:fa_up} show the profiles of the downward and upward availability factors, respectively.

In general, downward availability factors depend on the amount of power being generated.
For example, PV generation factors are close to the power generation curve (the same happens for wind generation factors).
For the BESS, instead, they exhibit high values during discharging and become zero when the BESS is charging at its nominal power.


Complementary to the case above, upward availability factors depend on the amount of available power.
Two patterns can be observed: one in which the factors remain close to zero, and another characterized by intermittent variations over specific time intervals.
Since the wind turbines at node 1 and the PV plant operate close to their rated capacity throughout the day, their corresponding upward availability factors remain close to zero.
The battery and wind farm at node 2, instead, have factors that virtually sum 1.
The intermittent intervals are imposed by the rolling horizon optimiser and mainly occur when one of the sources operates near its maximum.
Availability is therefore present at some instants and absent at others.
This intermittency is not reflected in the output power.
Indeed, availability factors are multiplied by $p^*_{SC,VPP}$ (see Fig.~\ref{fig:Secondary_Controller}), which is usually small compared with the total VPP  power output.
A possible approach to mitigate the intermittency is to incorporate moderate penalty terms.
\vspace{-0.2cm}
\subsection{Case 2: Operation Under Forecast Uncertainty}
\begin{figure}[!t]
\centering
\vspace{-0cm}
\includegraphics[width=0.99\columnwidth]{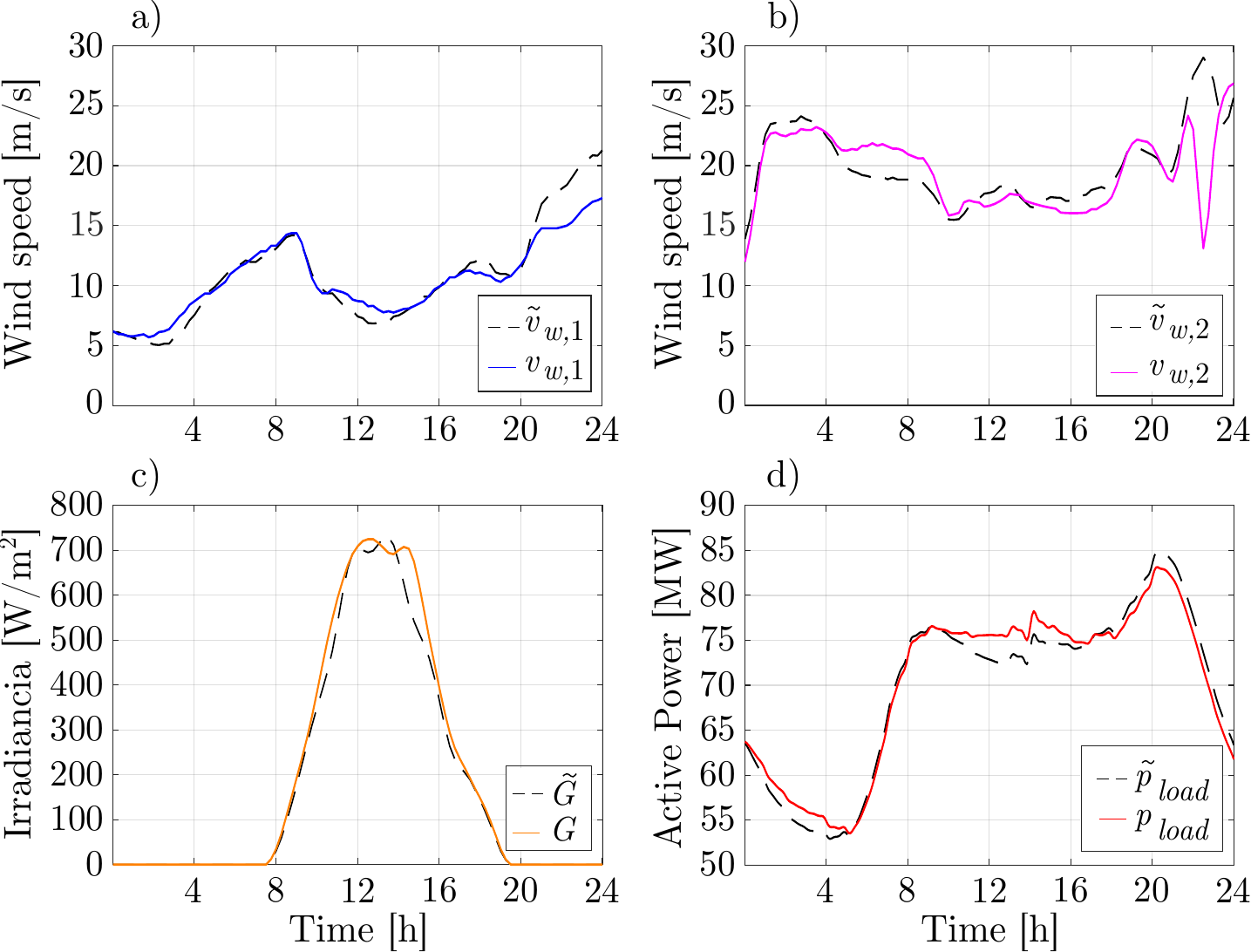}
\vspace{-0.5cm}
\caption{Parameters of the RESs and VPP load: a) Wind speed at node~1; b) Wind speed at node~2; c) Irradiance; d) VPP load. }
\vspace{-0.3cm}
\label{fig:Resources}
\end{figure}

\begin{figure}[!t]
\centering
\vspace{-0.2cm}
\includegraphics[width=0.94\columnwidth]{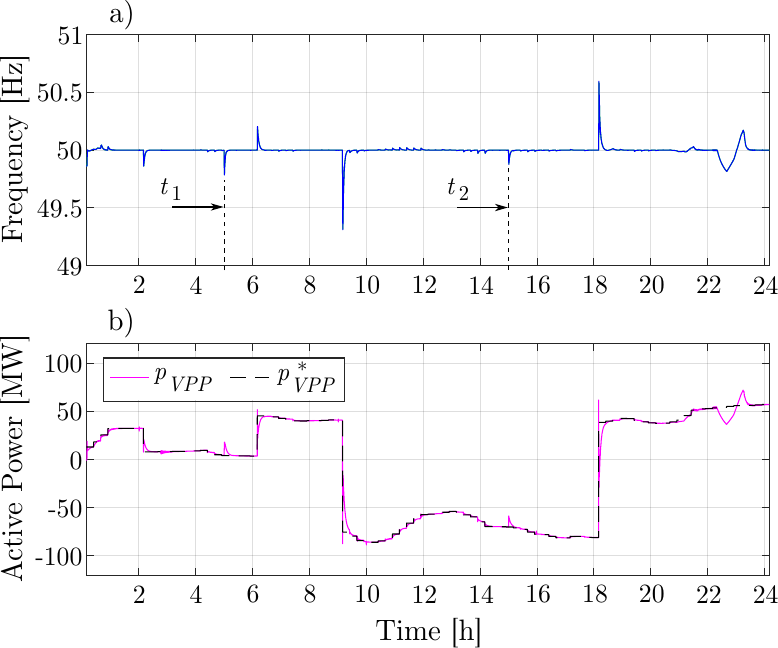}
\vspace{-0.4cm}
\caption{Time response of the power system under forecast uncertainty: a)~frequency; b)~active power injected by the VPP.}
\vspace{-0.2cm}
\label{fig:Freq_and_P_vpp_C2_2}
\end{figure}

\begin{figure}[!t]
\centering
\vspace{-0.1cm}
\includegraphics[width=0.99\columnwidth]{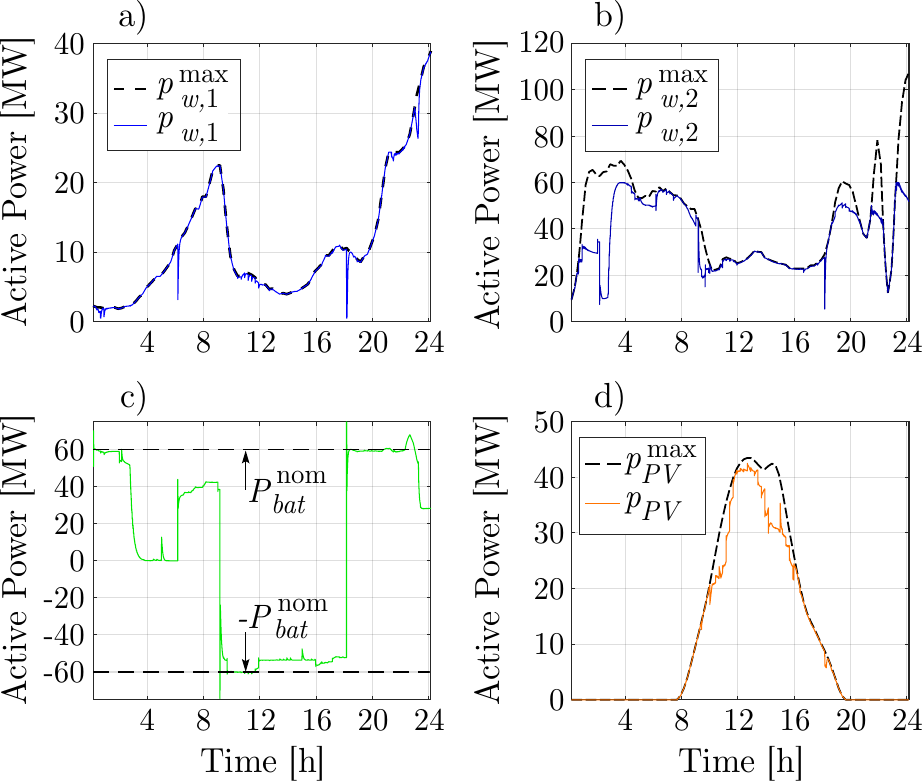}
\vspace{-0.5cm}
\caption{Active power profiles injected by the single VPP units under forecast uncertainty: a) wind farm~1; b) wind farm~2; c) PV plant; c) BESS.}
\vspace{-0.25cm}
\label{fig:P_VPP_units_C2}
\end{figure}

In this case, it is considered that the actual values of RES generation and system load deviate from the forecasted values employed in the DAM, as illustrated in Fig.~\ref{fig:Resources}.
\paragraph{Frequency and VPP Active Power Response}
The network frequency and the VPP output power are presented in Fig.~\ref{fig:Freq_and_P_vpp_C2_2}. 
As in Case 1, it can be observed that, following the transients, these signals return to their reference values.
Responses similar to those observed in Case 1 were obtained even when the actual operating values were considered.
In the final interval of the day, from $t \approx$ 22.5~h, there is a clear difference with respect to Case 1, where frequency and VPP power output deviate from their references.
This behaviour is mainly attributed to the significant difference between the forecast and actual resources, particularly the much smaller wind generation with respect to the forecast, as illustrated in Fig.~\ref{fig:Resources}(a) and Fig.~\ref{fig:Resources}(b).
This produces an undergeneration that the BESS cannot compensate for, since it is already operating at its nominal power, and it is compensated only when the wind speed at node 2 rises again and the wind farm recovers.
\vspace{-0.2cm}
\paragraph{Generation and Storage of VPP Units}
Fig.~\ref{fig:P_VPP_units_C2} shows the active powers injected by the VPP generators and storage units.
Compared to Case 1, the wind farm connected to node 2 and the PV plant operate below 100~\% of their rated capacity for a longer period, as illustrated in Fig.\ref{fig:P_VPP_units_C2}(b) and Fig.\ref{fig:P_VPP_units_C2}(d), respectively.
This difference is particularly noticeable in the PV plant, mainly in the hours in which there is more generation than expected, at around $t=$10~h and $t=$14-16~h.
Regarding the wind farm at node 1, it operates at 100~\% of its rated capacity the whole day. 
However, unlike in Case 1, it does not reach saturation.
Finally, as shown in Fig.~\ref{fig:P_VPP_units_C2}(c), the BESS profile in this case deviates more from the DAM schedule to compensate for the differences in RES generation.


\vspace{-0.2cm}
\section{Conclusion}
\label{sec.conclusion}
This work presented a centralised VPP controller based on a rolling-horizon optimisation problem. 
The VPP is used to provide energy and reserve capacity for automatic frequency restoration and its controller aims to closely follow the commitments established in the day-ahead market.
Nevertheless, the key aspect of the proposed controller is that it considers both the technical performance and the economic aspects of the individual VPP units. 
The distribution of the setpoints between the units is assigned according to resource availability and operational constraints, in such way that the operational cost of the entire VPP is minimised.
By combining the economic scheduling with the real-time allocation between units, the proposed controller aims to bridge the gap between the optimisation-oriented and the control-oriented approaches.

The main developments were validated using a power system that includes detailed dynamic models of the VPP elements.
To confirm the algorithm performance, the proposed rolling-horizon optimiser was tested under ideal forecast conditions and under presence of forecast uncertainty.
In both cases, the frequency was restored after deviations were introduced by step changes of either the VPP power reference or the total system load.
The VPP active power followed its reference value committed in the DAM optimisation, and the rolling-horizon optimiser effectively distributed the burden between units based on resource availability and technical constraints.
The results validated the effectiveness of the proposed controller.

Regarding the allocation of availability factors, the results showed that downward availability factors exhibit more stable dynamics compared to upward factors.
This is because downward factors depend on the power being generated, which varies smoothly, whereas upward factors depend on the unused available power.
The latter can become intermittent when sources operate close to their maximum. 
This intermittency, however, is generally not reflected in the power output of the VPP because it affects only a small part of it, more precisely, the VPP contribution to frequency restoration.
The controller maximises the utilisation of renewable generators, operating them close to their maximum capacity, except when there are technical constraints, such as excessively high wind speeds. 
Nonetheless, renewable sources provide upward reserves when required to comply with reserve commitments, or when this option is more profitable.
In the tests, the role of BESS was to compensate forecast errors adjusting its schedule according to the deviations in the renewable generation, thus allowing the VPP to follow its power reference.

Future work will extend the proposed algorithm to consider the power flows within the VPP and related losses, as well as internal constraints such as node voltages and line loading.

\vspace{-0.15cm}
\bibliography{Reference}

\begin{thebibliography}{10}

\bibitem{VPPs_Models_and_Markets}
N.~Naval and J.~M. Yusta, ``Virtual power plant models and electricity markets-a review,'' {\em Renew. Sustain. Energy Rev.}, vol.~149, p.~111393, 2021.

\bibitem{ZARE2026116448}
A.~Zare, M.~Shafie-khah, P.~Siano, and G.~C. Lazaroiu, ``A systematic review of virtual power plant configurations and their interaction with electricity, carbon, and flexibility markets,'' {\em Renew. Sustain. Energy Rev.}, vol.~226, p.~116448, 2026.

\bibitem{GHOLAMI2025101959}
K.~Gholami, M.~T. Arif, M.~E. Haque, A.~Arefi, and S.~Muyeen, ``Comprehensive review of cutting-edge virtual power plant advancements for flexibility enhancement in future power grids,'' {\em Energy Strategy Rev.}, vol.~62, p.~101959, 2025.

\bibitem{VPPs_concepts}
J.~F. Venegas-Zarama, J.~I. Muñoz-Hernandez, L.~Baringo, P.~Diaz-Cachinero, and I.~De~Domingo-Mondejar, ``A review of the evolution and main roles of virtual power plants as key stakeholders in power systems,'' {\em IEEE Access}, vol.~10, pp.~47937--47964, 2022.

\bibitem{VPP_DAM}
A.~Baringo, L.~Baringo, and J.~M. Arroyo, ``Day-ahead self-scheduling of a virtual power plant in energy and reserve electricity markets under uncertainty,'' {\em IEEE Trans. Power Syst.}, vol.~34, no.~3, pp.~1881--1894, 2019.

\bibitem{VPP_Robust_Scheduling}
Y.~Zhang, F.~Liu, Z.~Wang, Y.~Su, W.~Wang, and S.~Feng, ``Robust scheduling of virtual power plant under exogenous and endogenous uncertainties,'' {\em IEEE Trans. Power Syst.}, vol.~37, no.~2, pp.~1311--1325, 2022.

\bibitem{Optimisation_Models_DAM}
D.~Fernández-Muñoz and J.~I. Pérez-Díaz, ``Optimisation models for the day-ahead energy and reserve self-scheduling of a hybrid wind-battery virtual power plant,'' {\em J. Energy Storage}, vol.~57, p.~106296, 2023.

\bibitem{Optimal_BESS_Participation}
B.~Xu, Y.~Shi, D.~S. Kirschen, and B.~Zhang, ``Optimal battery participation in frequency regulation markets,'' {\em IEEE Trans. Power Syst.}, vol.~33, no.~6, pp.~6715--6725, 2018.

\bibitem{DVPP_Control_Design}
V.~Häberle, M.~W. Fisher, E.~Prieto-Araujo, and F.~Dörfler, ``Control design of dynamic virtual power plants: An adaptive divide-and-conquer approach,'' {\em IEEE Trans. Power Syst.}, vol.~37, no.~5, pp.~4040--4053, 2022.

\bibitem{Direct_Part_DVPP_in_FC}
M.~E. Adabi and B.~Marinescu, ``Direct participation of dynamic virtual power plants secondary frequency control,'' {\em Energies}, vol.~15, no.~8, 2022.

\bibitem{Load_Freq_Control_Wang}
Z.~Wang, Y.~Wang, and L.~Xie, ``Load frequency control of multiarea power systems with virtual power plants,'' {\em Energies}, vol.~17, no.~15, 2024.

\bibitem{GUO202593}
J.~Guo, C.~Dou, D.~Yue, B.~Zhang, Z.~Zhang, and Z.~Zhang, ``Coordinated power control of virtual power plant for frequency regulation considering cyber-physical uncertainties,'' {\em Cyber-Phys. Energy Syst.}, vol.~1, no.~1, pp.~93--103, 2025.

\bibitem{Cord_Freq_Reg_Oshnoei}
A.~Oshnoei, M.~Kheradmandi, F.~Blaabjerg, N.~D. Hatziargyriou, S.~Muyeen, and A.~Anvari-Moghaddam, ``Coordinated control scheme for provision of frequency regulation service by virtual power plants,'' {\em Appl. Energy}, vol.~325, p.~119734, 2022.

\bibitem{DVPP_Robust_Freq_regul}
X.~Zhu, H.~Geng, H.~Qing, G.~Ruan, and X.~He, ``Dynamic virtual power plants with robust frequency regulation capability,'' {\em IEEE Trans. Ind. Appl.}, vol.~62, no.~1, pp.~132--143, 2026.

\bibitem{Min_Reserve_and_Allocation_Zhu}
X.~Zhu, G.~Ruan, and H.~Geng, ``Optimal frequency support from virtual power plants: Minimal reserve and allocation,'' {\em Appl. Energy}, vol.~392, p.~125913, 2025.

\bibitem{Optimal_VPP}
A.~Bolzoni, A.~Parisio, R.~Todd, and A.~J. Forsyth, ``Optimal virtual power plant management for multiple grid support services,'' {\em IEEE Trans. Energy Convers.}, vol.~36, no.~2, pp.~1479--1490, 2021.

\bibitem{MPC_Amini}
M.~Amini, A.~Khurram, A.~Klem, M.~Almassalkhi, and P.~D. Hines, ``A model-predictive control method for coordinating virtual power plants and packetized resources, with hardware-in-the-loop validation,'' in {\em IEEE Power \& Energy Society General Meeting (PESGM)}, pp.~1--5, 2019.

\bibitem{Gulotta}
F.~Gulotta, P.~Crespo~del Granado, P.~Pisciella, D.~Siface, and D.~Falabretti, ``Short-term uncertainty in the dispatch of energy resources for {VPP}: A novel rolling horizon model based on stochastic programming,'' {\em Int. J. Electr. Power Energy Syst.}, vol.~153, p.~109355, 2023.

\bibitem{SFC_of_Dist_Level_VPP_Zhuang}
Z.~Zhuang, X.~Dou, and L.~Yu, ``Secondary frequency control strategy of distribution-level virtual power plant,'' in {\em 2025 7th Asia Energy and Electrical Engineering Symposium (AEEES)}, pp.~645--650, 2025.

\bibitem{Real_time_op_VPPs_Chen}
Q.~Chen, R.~Lyu, H.~Guo, and X.~Su, ``Real-time operation strategy of virtual power plants with optimal power disaggregation among heterogeneous resources,'' {\em Appl. Energy}, vol.~361, p.~122876, 2024.

\bibitem{Decentralized_FR_for_VPPs}
X.~Sun, H.~Xie, D.~Qiu, Y.~Xiao, Z.~Bie, and G.~Strbac, ``Decentralized frequency regulation service provision for virtual power plants: A best response potential game approach,'' {\em Appl. Energy}, vol.~352, p.~121987, 2023.

\bibitem{Enabling_DER_Freq_reg_Markets}
P.~Srivastava, C.-Y. Chang, and J.~Cortés, ``Enabling {DER} participation in frequency regulation markets,'' {\em IEEE Trans. Control Syst. Technol.}, vol.~30, no.~6, pp.~2391--2405, 2022.

\bibitem{10764748}
Y.~Deng, X.~Fang, N.~Gao, and J.~Tan, ``Multi-timescale modeling framework of hybrid power plants providing secondary frequency regulation,'' {\em IEEE Open Access J. Power Energy}, vol.~11, pp.~595--609, 2024.

\bibitem{Kundur_book1}
P.~Kundur, {\em Power System Stability and Control}.
\newblock McGraw-Hill, 1994.

\bibitem{Marco_IECON_2025}
M.~V. Avendaño-Caiza, J.~D. Rios-Peñaloza, J.~Roldán-Perez, J.~L. Rodríguez-Amenedo, and M.~Prodanović, ``Considering resource availability in control of dynamic virtual power plant used for provision of frequency restoration reserve,'' in {\em Annual Conference of the IEEE Industrial Electronics Society (IECON)}, pp.~1--6, 2025.

\bibitem{AccessModels}
M.~V. Avendaño-Caiza, J.~Roldán-Pérez, N.~Jankovic, J.~L. Rodríguez-Amenedo, J.~D. Rios-Peñaloza, and M.~Prodanović, ``Application of model order reduction in evaluation of small-signal stability of power systems with virtual power plants,'' {\em IEEE Access}, vol.~14, pp.~44277--44292, 2026.

\bibitem{P_Systems_Dynamics}
M.~Eremia and M.~Shahidehpour, {\em Handbook of Electrical Power System Dynamics: Modeling, Stability, and Control}.
\newblock John Wiley, 2013.

\bibitem{Cp_calculate}
M.~Carpintero-Rentería, D.~Santos-Martín, A.~Lent, and C.~Ramos, ``Wind turbine power coefficient models based on neural networks and polynomial fitting,'' {\em IET Renew. Power Gener.}, vol.~14, pp.~1841--1849, 2020.

\bibitem{PV_model1}
S.~Hara, H.~Douzono, M.~Imamura, and T.~Yoshioka, ``Estimation of photovoltaic cell parameters using measurement data of photovoltaic module string currents and voltages,'' {\em IEEE J. Photovolt.}, vol.~12, no.~2, pp.~540--545, 2022.

\bibitem{Batteries_mod2}
L.~Gao, S.~Liu, and R.~Dougal, ``Dynamic lithium-ion battery model for system simulation,'' {\em IEEE Trans. Compon. Packag. Technol.}, vol.~25, no.~3, pp.~495--505, 2002.

\bibitem{VSC_book}
A.~Yazdani and R.~Iravani, {\em Voltage-Sourced Converters in Power Systems: Modeling, Control, and Applications}.
\newblock John Wiley, 1~ed., 2010.

\bibitem{Z_virtual_Jav}
A.~Rodríguez-Cabero, J.~Roldán-Pérez, and M.~Prodanovic, ``Virtual impedance design considerations for virtual synchronous machines in weak grids,'' {\em IEEE J. Emerg. Sel. Topics Power Electron.}, vol.~8, no.~2, pp.~1477--1489, 2020.

\bibitem{BOE_137_250424}
{CNMC}, ``{Resolution of 25 April 2024 on balancing terms and conditions and operating procedures for participation in the MARI and PICASSO platforms}.'' Bolet\'in Oficial del Estado, no. 137, pp. 66281--66524, 2024.
\newblock BOE-A-2024-11535.

\bibitem{Diego_TSTE_Central}
J.~D. Rios-Peñaloza, A.~Prevedi, M.~Prodanović, and J.~Roldán-Pérez, ``Central controller for photovoltaic power plants with hybrid energy storage systems providing automatic frequency restoration reserve,'' {\em IEEE Trans. Sust. Energy}, vol.~17, no.~3, pp.~2543--2557, 2026.

\bibitem{Yalmip}
J.~L{\"{o}}fberg, ``{YALMIP}: A toolbox for modeling and optimization in {MATLAB},'' in {\em Proc. CACSD Conf.}, (Taipei, Taiwan), 2004.

\end{thebibliography}
\bibliographystyle{ieeetr}

\end{document}